# Spall Craters in the Solar System

Keith A. Holsapple

P.O. Box 305

Medina, WA 98039

kholsapple0@gmail.com

Kevin R. Housen

5002 Varco Rd NE

Tacoma, WA 98422

kbhousen@comcast.net

Manuscript has 20 pages, including 11 Figures and 2 Tables.

# Abstract

Small high-speed impact craters formed in rocks, ice, and other brittle materials consist of an outer, broad shallow concentric region formed by tensile fracture (spall), surrounding a smaller central "pit" crater of greater depth. On the Earth, that "spall crater" morphology ceases to exist for craters greater than a few meters in diameter. They are not commonly recognized for craters in the solar system but might be an issue for cratering on the small brittle asteroids.

We consider the physics of the processes of shock-wave spall cratering and formulate the scaling laws to apply those processes to the bodies of the solar system. Our scaling is based upon analyses of shock-wave propagation and tensile fracture mechanisms, including the important feature of size-dependent tensile fracture, and the role of gravity in lofting spalled material to form the outer parts of the spall craters. We consider the existing scaling laws for cratering in the strength regime and derive the conditions for which spall features will be present or absent. The conditions giving rise to spall cratering are found to be a distinct subset of the 'strength' regime, forming a new sub-regime of cratering.

We find that this regime may be very consequential for planetary cratering; in fact, it might dominate all cratering on small rocky asteroids. That has important implications in the interpretation of crater counts and the expected surface effects for rocky, 10-100 km objects.

# 1. Introduction

It is well documented that laboratory-scale craters in rocks and ices are different than large-scale craters in rocks. In the lab, small-scale craters in rocks and other brittle materials typically have a central deeper "pit" crater surrounded by a shallow broad region, formed by the process of spallation, in which the surface material fails in tension and is ejected upwards from the surface. There is no raised rim on the outer spall crater. Often the material from that spall region can be found in broken fragments (blocks) near the crater.

Figures 1 through 5 show examples of small, mm to cm-scale spall craters formed in rock (1-4) and glass (5). These figures all show the same distinctive features of spall craters: a broad shallow rimless region that is typically 3 to 4 times larger in diameter than a central pit crater (e.g. Kawakami et al., 1983; Hörz et al., 1971).

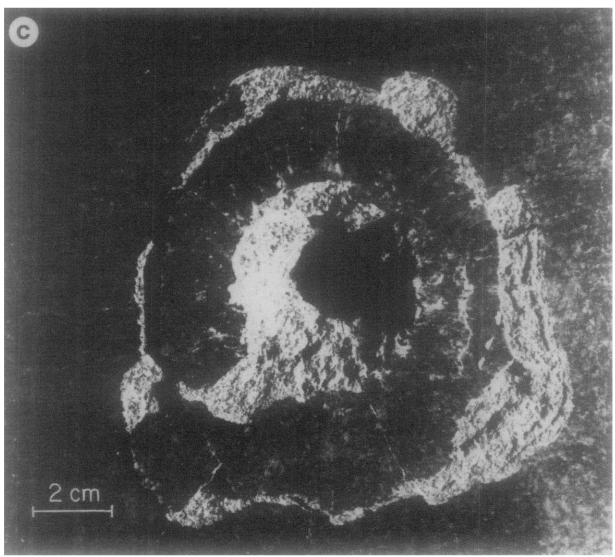

Figure 1. A cm-sized crater formed in San Marcos Gabbro from an impact at 4.6 km/s. The central bowl-shaped crater is surrounded by a broad flat region where the surface layer was removed after a tensile failure (Lange *et al*. 1984).

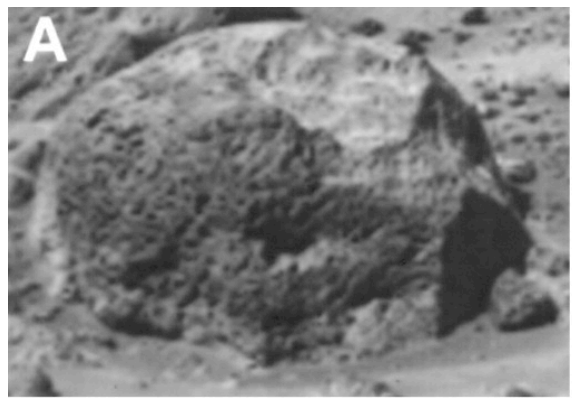 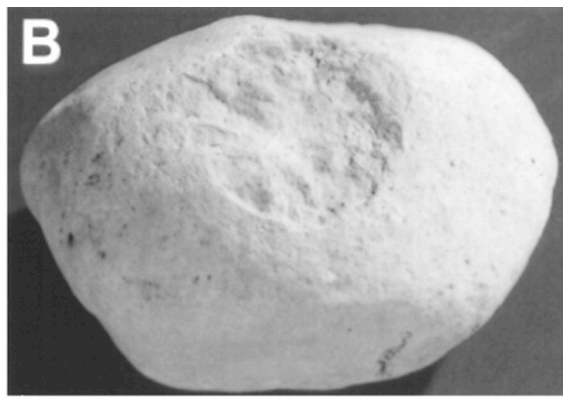

Figure 2. On the left is a 25 cm spall crater in a rock photographed in the Pathfinder mission to Mars. On the right is a terrestrial laboratory simulation of a 10 cm diameter crater created by the impact of a 1/8" glass sphere at 6.0 km s$^{-1}$ into a limestone boulder. (Hörz et al., 1999).

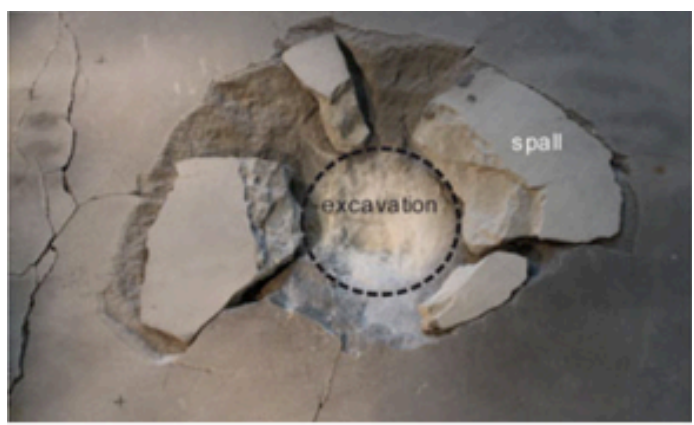

Figure 3. A spall crater with the spall fragments retrieved from nearby the crater and re-inserted into the target after the formation. Characteristically there are a few large fragments and also a large number of small fragments from the central excavation bowl. (from Dufresne et al., 2013)

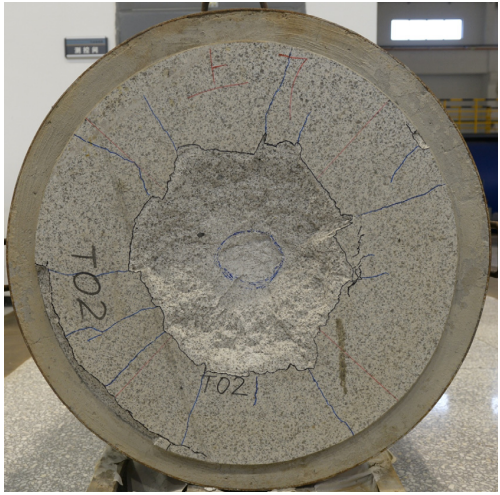

Figure 4. A granite target impact at 3.55 km/s, one in a series if 15 experiments. The outer spall diameter is 19 times the projectile diameter and about 4 times the inner bowl diameter. Wang et al. 2020 Appl. Sci. 2020, 10,1393.

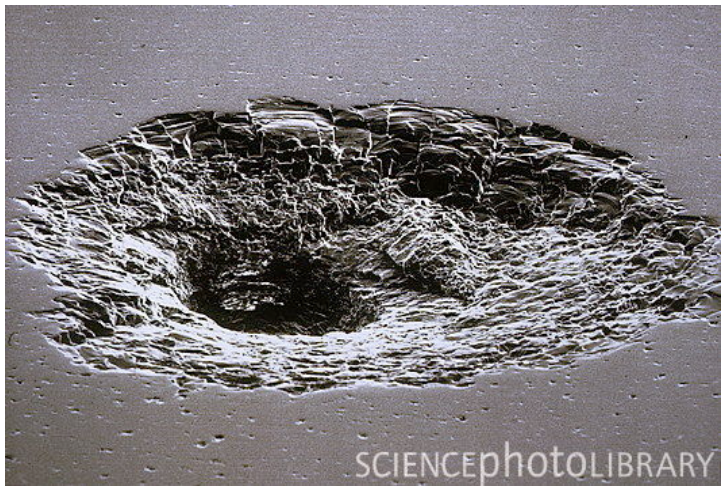

Figure 5. A Scanning Electron Micrograph (SEM) of a mm-sized crater in the surface of a window of Space Shuttle Challenger. The crater is about 640 microns in diameter and 630 microns in depth, surrounded by a spall zone of 2.4 mm diameter (Hörz et al., 1971).

r than a few to several meters, that broad spall region ingle bowl-shaped crater. An example, Fig. 6 shows the rface explosion in rock.

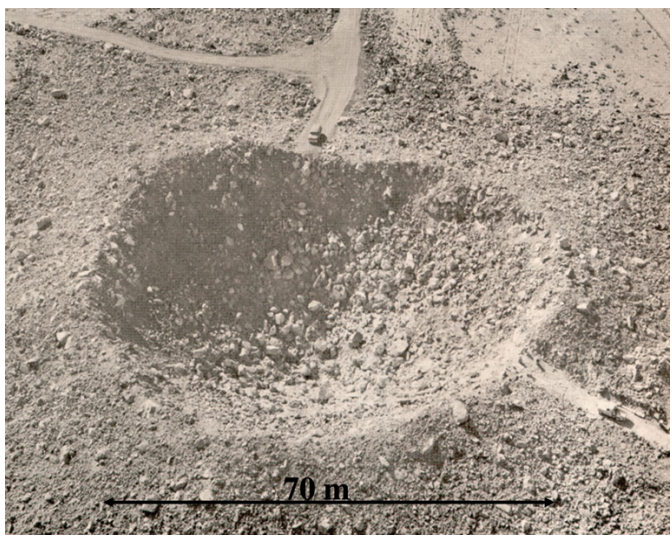

Figure 6. A 70 m diameter crater formed by a high explosive charge detonated in rock. Note the raised rim and absence of large fragments. This is not a spall crater. Spall craters have spall fragments with dimensions on the order of a significant part of the crater diameter and no raised rim.

Interestingly, about fifty years ago, Gault, 1973 made the following observation regarding craters formed on Earth in rocks by high explosives: "In the range of $10^{15}$ to $10^{16}$ ergs, large spall-like plates of rock around the periphery of the craters, which at smaller scale are ejected from the craters, are apparently heaved upwards with a very low velocity only to settle down subsequently with small upward final displacements". He further noted: "It is probable that in gravitational fields significantly lower that one G (i.e. lunar and asteroidal), the large spall plates will eject from the crater." Using his scaling, those energy levels would correspond to crater diameters of 5 to 9 m.

This process envisioned by Gault (1973) has been illustrated in impact experiments performed by the authors. The experiments were conducted on a geotechnical centrifuge, which allows large-scale craters to be simulated at small scale by effectively increasing gravity (Schmidt and Holsapple 1980). In this method, small experiments at high acceleration simulate those at large scale[1]. For example, an experiment conducted at 10x terrestrial gravity (10G) simulates an impact that is 10x larger than an otherwise identical experiment conducted at 1G. Figure 7 shows the results of 1.5 km/s of aluminum projectiles onto granite targets. The image on the left shows a roughly 2 cm pit crater surrounded by a larger, roughly 7 cm, spall zone[2]. The middle image shows a crater formed under 10G acceleration. High speed video of the events showed that some of the peripheral spall plates were ejected but were retained within the crater. At the highest acceleration of 500G, simulating the formation of a 10m central pit crater at 1G, the ejection of the surrounding spall plates was not sufficient to dislodge them. Black arrows indicate some of those retained plates.

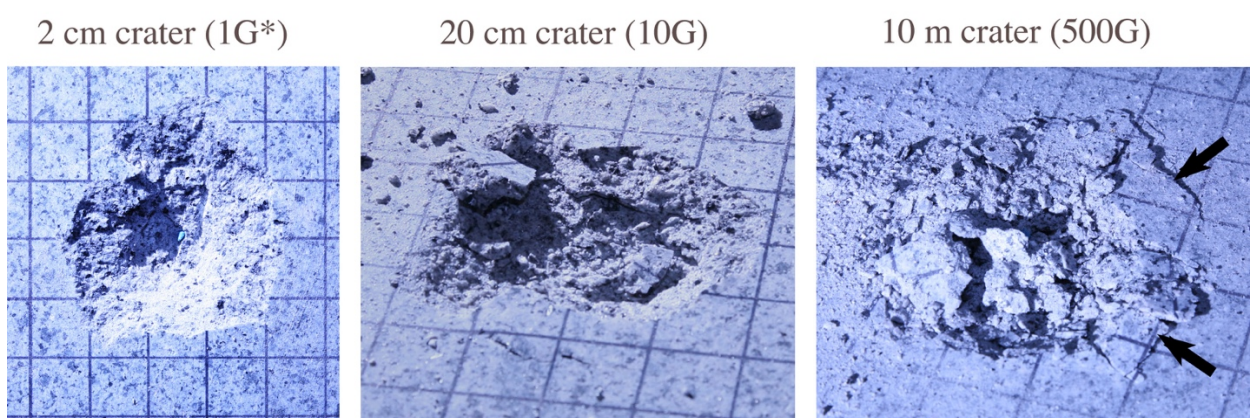

Fig. 7. Experimental craters formed in High Cascade Granite targets on a geotechnical centrifuge. Projectiles were ~3 gm aluminum cylinders impacting at 1.5 km/s. Higher accelerations simulate larger events. At the highest acceleration tested (500G), the ejection velocities of the peripheral spall plates (marked by arrows) were not sufficient to dislodge them. Gridlines are spaced 2 cm apart.

The formation and possible ejection of spall plates creates a problem when trying to simulate large cratering events with small-scale experiments in the laboratory in these brittle materials. Should one, for example, include the volume of the spalled material in scaling formulas? Or instead, should one ignore that volume and only consider the volume of the central crater region? What is the fate of the material in that spalled region? How do we relate the small-scale results to the larger events of interest? To date, small-scale spall craters have often been treated as an annoyance that precluded making meaningful small laboratory experiments of cratering in brittle materials.

The existence of spall craters and how they scale with increasing event size is the subject of this paper. We speculate that these spall craters are very important in the overall picture of asteroid cratering; and show the existence, in addition to the standard strength and gravity regimes, of an additional important type of crater formation. We develop the scaling theory for those spall craters.

---

[1] With the provision that the centrifuge experiment does not replicate the lower material strength expected for a large crater – as discussed in the next section.
[2] Note, an experiment was not actually performed at 1G. Instead, this was simulated by inverting the 500G crater to allow the retained debris to fall out of the crater, thereby simulating the escape of the debris that would have occurred at 1G.

## 2. Scaling theory I: Bowl craters in the strength regime

We begin with the now standard scaling arguments (Holsapple and Schmidt 1987, Housen, Schmidt, and Holsapple 1983, Holsapple 1993) for an impact into a fixed, strength dominated (smaller craters), brittle material. Any characteristic crater dimension such as the outer crater diameter $D$ is some function of the conditions of that impact. Specifically, it is determined by the impactor radius $a$, the impactor velocity $U$, the impactor mass density $\delta$, the target mass density $\rho$, and some measure $Y$ of the target shear strength:

$$D = F[a, U, \delta, \rho, Y] \tag{1}$$

This defines what is called the 'strength regime' of cratering. It is assumed that there is no effect of the target surface gravity $g$ (the 'gravity regime'). This case applies for small craters in cohesive materials such as rock.

At high impact speeds $U$, and for crater size scales significantly larger than the impactor radius $a$, there are not separate dependences on $a$, $U$, $\delta$; but instead only a single combined "coupling parameter" measure $C$ that must be of the power law form $C = a\, U^\mu \delta^\nu$. That is the central idea of the existence of a point-source measure: a point source measure cannot have separate time or length scales. The validity of this point-source assumption for hypervelocity impacts has been well proven by theory and experiments over the last several decades. For rocks, the exponent $\mu$ is found to be about 0.55, and $\nu$ is about 0.4 (e.g. Holsapple, 1993).

For a point source Eq. (1) simplifies to:

$$D = F[aU^\mu \delta^\nu, \rho, Y] \tag{1}$$

There are only four variables in this equation and three units of measurement (mass, length, and time) so, according to the theory of dimensional analysis, the four variables must combine into one non-dimensional group, as follows

$$\frac{D}{a}\left[\frac{Y}{\rho U^2}\right]^{\frac{\mu}{2}} \left[\frac{\rho}{\delta}\right]^\nu = K_D \tag{3}$$

where $K_D$ is a fixed constant. Note that the central feature of this form is that, when expanded out, $a$, $U$ and $\delta$ occur only in the power form in (2) required of the coupling parameter. It also shows that, when impacts have fixed target and impactor material, velocity $U$ and target strength $Y$, the radius $D$ of craters will be proportional to the impactor radius $a$. The term with the ratio of target to impactor mass densities is close to unity, so it will be ignored.

For impacts at an angle ∅ measured from the perpendicular direction, a standard assumption is that the result will be the same as for a perpendicular impact at velocity $U \cos ∅$. as a result, the eq. (3) gives that

$$D = K_D a \left[ \frac{Y}{\rho (U \cos ∅)^2} \right]^{\frac{-\mu}{2}} \tag{4}$$

The scaling defined by this equation is called "cube root" scaling, a term introduced long ago in studies of explosive cratering. At a fixed velocity, it has crater linear dimensions proportional to the cube-root of the impactor energy (which is proportional to $a^3$), thereby being proportional to the linear dimension of the impactor. Therefore, the crater volume is proportional to the impactor mass and energy. Another feature of cube-root scaling is that the pressure, stress tensor and velocity associated with the outgoing shock will be identical for two impacts at the same impact velocity and at the same homologous (scaled to $a$) locations (e.g. Holsapple 1993).

But a further circumstance for cratering in brittle materials, like rock, is that the material strength likely depends on size scale. The strength of a brittle material is determined by an internal distribution of small cracks and flaws. (e.g. Grady and Kipp, 1980). That physical feature leads to an effective strength that is less for a large event than for a small one, a concept that has been long recognized both in impact cratering (Gault, 1973) and general geomechanics. The strength decreases with the crater scale whenever the scale is greater than some length $D_0$, a common form is $Y = Y_0(D_0/D)^{1/n}$ for some value $n$. A nominal value of $n$ is about 4 although a "fully cracked" material has the exponent $n=2$ (Holsapple and Housen, 1986; Housen and Holsapple,1999; Holsapple, 2009).

There is, of course, more than one kind of "strength" (e.g. Holsapple, 2009). Bowl craters are primarily determined by a shear strength (technically called 'cohesion'), while spallation is determined by a tensile strength. Tensile strength likely has a stronger size dependence than shear strength. Therefore, for the formation of bowl craters we assume $n=n_1=4$. For spallation, we use $n=n_2=2$.

Suppose that $Y = Y_0(D_0/D)^{1/n}$ for some value $n$. We insert that into (3) and then solve for a bowl crater radius $D_B$ to get its dependence on the impactor radius:

$$\frac{D_B}{a} = K_{DB} \left( \left( \frac{a}{D_0} \right) \left( \frac{Y_0}{\rho (U \cos ∅)^2} \right)^{-n} \right)^{\mu/(2n-\mu)} \tag{5}$$

Now, for a fixed target and velocity, the scaled crater radius $D_B/a$ increases with the power $\mu/(2n - \mu)$ of the impactor radius. With values $\mu = 0.55$ and $n=4$ that is a power of 0.07. Alternatively, $D_B \sim a^{2n_1/(2n_1-\mu)}$ where the exponent would be 1.07, a value larger than the linear dependence on impactor dimension of cube-root scaling. There is more size dependence if $n=2$, for which the scaled diameter increases with impactor size to the power of 0.16. The scaling app at Holsapple, 2022 has the leading coefficient $K_{DB}=1.15$.

This form would apply whenever the strength is distinctly greater than the lithostatic compressive force due to gravity, i.e:

$$Y = Y_0(D_0/D)^{1/n_1} > \rho g D, \tag{6}$$

The crater diameters at that transition are given as

$$D = D_{strength} = D_0 \left( \frac{Y_0}{\rho g D_0} \right)^{\frac{n_1}{n_1+1}} \tag{7}$$

Of course, that strength-gravity transition is broad, this diameter can be considered the center of it.

Eq (5) can be put into another useful and simple form for scaling from one event with impactor dimension $a_1$ and bowl crater diameter $D_{B1}$ to another in the same target and at the same impact velocity for impactor $a_2$ and crater $D_{B2}$ :

$$\frac{D_{B2}}{D_{B1}} = \left(\frac{a_2}{a_1}\right)^{2n_1/(2n_1-\mu)} \qquad (8)$$

In the (unrealistic) case where strength is not size dependent, then $n_1=\infty$, and this is just the cube-root scaling from one event to another in a fixed material, with the crater size proportional to impactor radius $a$.

At fixed velocity (only), the impactor radius is proportional to the cube-root of the impactor energy. Thus, (8), with $n_1 = 4$ and $\mu = 0.55$ gives the crater size increasing with a power of impact energy of 0.386. This is consistent with measurements; for example Burchell et al., 1998 in solid $CO_2$ targets reported powers on energy of 0.40 for depth and 0.38 for radius, or 1.2 and 1.14 on impactor radius $a$. Gault, 1973 reported an exponent of 1.13 for the volumes, implying a power of 0.38 on energy for a linear dimension. Those experimental results support the concept of a size-dependent strength.

## 3. Scaling theory II: Spall craters

The introduction described the general features of a spall crater. The sizes of spall craters, assuming again the point-source measure, also are given by the same results as above, but with a smaller exponent $n_2$ for the spall strength instead of $n_1$ for a shear strength (larger size dependence), and a different leading constant. We omit the details. The result is given by

$$\frac{D_S}{a} = K_S \left(\left(\frac{a}{D_0}\right)\left(\frac{Y_0}{\rho(U\cos\emptyset)^2}\right)^{-n_2}\right)^{\mu/(2n_2-\mu)}, \qquad (9)$$

the same as for a bowl crater, but with the size exponent $n_2$ governing spall strength and a new coefficient $K_S$. Further, the strength $Y_0$ used here would be a tensile spall strength rather than a shear strength, although they are of the same order.

### *3.1. Numerical Example*

As the dominant example, we will use the well-documented measurements for lab spall craters in San Marcos Gabbro given by Polanskey and Ahrens, 1990. Several experiments used 1/8" (3.2 mm) diameter spherical impactors at a velocities of 4.6 to 6.49 km/s. The craters were like those shown in Figs. 3 to 6 above. The outer spall craters had diameters from 6.5 to 10.5 cm and inner diameters of about 2.5 cm diameter. After the impact, they recovered 14 distinct surface spall 'plate' fragments, with thicknesses ranging from 2 to 9 mm. The measured initial vertical spall velocities ranged from 3 to 30 m/sec. The measure tensile strength was $Y_0=1.5\ 10^9$ dynes/cm², and the target density was 2.9 g/cm³.

Their specific shot 840901 used an aluminum projectile with radius $a=0.159$ cm, velocity $U=6.49$ km/s, and produced a spall crater diameter of 6.5 cm. Those results are compatible with the strength scaling of (9) with $\mu = 0.55$, $D_0$=10 cm, $n_2$=2 and the leading constant as $K_S$=9.34. The crater diameter in this case is a factor of 2.5 larger than for a simple bowl crater.

What about the spall fragments? Consider the typical case of a spall fragment of thickness 5 mm with a typical initial vertical speed of 20 m/s. If free to do so, on earth (ignoring atmospheric drag) it would be cast to a height

$$h = \frac{v^2}{2g}, \qquad (10)$$

or 20 m, about 4000 times its thickness, and it would be entirely removed from the crater region. For a hypothetical larger impactor, what would the dimensions and spall velocity be? The answers to that question are the basis for the scaling.

A characteristic ejecta or spall velocity of a material element in a strength-determined cratering event at a given scaled range $r/a$ is (Housen et al., 1983):

$$v = k_v \sqrt{\frac{Y}{\rho}}. \qquad (11)$$

Therefore, if the strength $Y$ decreases with size, so will the velocity imparted to a spall region. Specifically, using the above form for the spall strength governed by an exponent $n_2$,

$$v = k_v \sqrt{\frac{Y_0}{\rho} \left(\frac{D_0}{D_S}\right)^{1/n_2}}. \qquad (12)$$

Using the Polanskey and Ahrens, 1990 average spall velocity of 20 m/s for calibration, the constant $k_v$~0.08.

The height to which it would be lofted, using (10) and (12) is

$$h = k_v^2 \frac{Y_0}{2g\rho} \left(\frac{D_0}{D_S}\right)^{1/n_2}. \qquad (13)$$

Spall craters will not be formed unless that height is greater than a few, say *mfac*~2 to 4 spall plate thicknesses $t$, which is some fraction $k_t$ of the spall crater diameter $D_S$

$$t = k_t D_S. \qquad (14)$$

From the Polansky and Ahrens experiments, $k_t$ is about 0.05. Therefore, spall plates will be lifted from the crater when

$$h = k_v^2 \frac{Y_0}{2g\rho} \left(\frac{D_0}{D_S}\right)^{1/n_2} > mfac\, k_t D_S, \qquad (15)$$

which can be solved for the minimum spall crater size for lofting of plates

$$D_S^* = D_0 \left(\frac{k_v^2\, Y_0}{2\, mfac\, \rho\, g\, k_t\, D_0}\right)^{n_2/(n_2+1)}. \qquad (16)$$

By using (9) the impactor radius $a^*$ creating this spall crater is

$$a^* = D_0\, (K_s)^{(\mu-2n_2)/(2n_2)} \left(\frac{k_v^2\, Y_0}{2\, mfac\, \rho g\, k_t\, D_0}\right)^{(2n_2-\mu)/(2+2n_2)} \left(\frac{Y_0}{\rho (U \cos \emptyset)^2}\right)^{\mu/2}. \qquad (17)$$

This spall crater regime is always a subset of the strength regime.

With *mfac*=3 and the constants listed below, this gives the spall crater upper limit in the non-dimensional form

$$D_S^* = 0.077\, D_0\, (Y_0/(D_0 \rho g))^{2/3}. \qquad (18)$$

as the largest spall diameter. Note that this result is independent of the scaling coefficient $\mu$ and the constant $K_s$ in (9). It would be formed from an impactor with the diameter $d^*$ proportional to the strength as

$$d^* \sim Y_0^{2+\frac{\mu}{3}}, \tag{19}$$

The various values for the constants implied by the Polanskey and Ahrens experiments are gathered in Table 1:

| Constant | Meaning | Value |
|---|---|---|
| $Y_0$ | Nominal Gabbro tensile strength | 1.5 $10^9$ dynes/cm$^2$ |
| $D_0$ | Specimen size having nominal strength | 10 cm |
| mu | Point source scaling coefficient for rocks | 0.55 |
| $n_2$ | Strength tensile size-exponent | 2 |
| Ks | Coefficient for spall crater diameter, Eq.(9) | 9.34 |
| $k_v$ | Coefficient for spall velocity, Eq.(11) | 0.08 |
| $k_t$ | Spall thickness/ crater diameter, Eq.(14) | 0.05 |
| mfac | Loft height per thickness | 3 |

Table 1. Constants that define the spall craters in Polanskey and Ahrens (1990).

From (18) a target with any finite strength will have some range of spall craters, although the range will become small as the strength approaches zero. For example, for an asteroid with diameter $D_{ast}$, and nominal mass density of 3.0 g/cm$^3$, for which $g = 4.19 \, 10^{-7} \, D_{ast}$, using (18), the largest spall crater on a 10 km body with the properties of Gabbro would be $D_S^* = 0.77 \left(\frac{1.5 \, 10^9}{3.0*10*0.419}\right)^{2/3} = 1.9$ km. On a 1 km body, all craters would be spall craters. However, if the strength were as low as $10^6$ (100 kPa), the largest spall crater on that 10 km target body would be only 14 m.

On Earth, an impact into a surface with the nominal Gabbro strength would be a spall crater if it were smaller than 11 m diameter, a result quite consistent with the Gault, 1973 observations for explosive craters. The largest spall crater's diameter is reduced by a factor of 5 for each decade of reduced strength.

These numerical results depend on the choices for the various constants defined above, which are not well determined, especially the target strength. A first-order indicator of that property for an asteroid is its mass density. Competent rocky materials will have mass densities close to the mass densities of individual minerals; on the order of 3 g/cm$^3$. A mass density substantially less than that indicates porosity and lower strength, which would reduce the probability of spall cratering. We should not expect the same cratering style nor crater scaling on the small asteroids Datura, Gaspra, Eros, Ida, compared to lower density bodies such as Ryugu, Itokawa, Bennu, our moon, Phobos, Deimos and others.

The primary difference from experiments on Earth is the level of the surface gravity $g$, which sets the height to which a spall plate is lofted. High gravity suppresses spall cratering because the spall plates are not permanently displaced. For an asteroid, the Eq. (18) notes that the crater size at which there is a transition from spall craters to non-spall craters scales with gravity to the power of 2/3. If the average transition size is about 10 meters on Earth, that transition size would include all craters on a small rocky asteroid. Specifically, with the mass density of 3 g/cm$^3$, the upper spall limit will exceed the asteroid diameter when D < 3 km. That is, <u>all craters formed on a solid rocky object smaller than about 3 km would be spall craters</u>. In addition, on such a small body, the spall fragments would escape the asteroid (see below).

## 3.2. Escaping Spall Fragments

By equating the spall velocity (Eq. 12) to an object's escape velocity, it is found that all spall fragments would escape from the surface for spall craters smaller than

$$D_S < D_0 \left(\frac{2\pi k_v^2 Y_0 G}{3g^2}\right)^{n_2}. \tag{20}$$

## 3.3. Fragment Sizes

The scaling has the spall thickness as some proportion of the spall diameter. In experiments, the spall fragments have average lateral dimensions a few times their thickness. For that reason, spall fragments would also scale with the crater diameter. The 10 cm craters of Polanskey and Ahrens produced about 10 fragments[3] each with lateral dimensions of a few to 10 cm. Then each 100 m spall crater on a low gravity asteroid (the largest possible from the Fig. 8 below) would produce 10 fragments (boulders or blocks) on the order of a few 10's of meters. One would not expect any fragments larger than 10's of meters. Only the very smallest would escape.

## 4. Application to Asteroids

These concepts and formulas can be tested against the observed craters on rocky asteroids and other bodies of the Solar System. There are two measures that are of particular importance. The first is simply the relation for the crater diameter as a function of the size of the impactor that created it. There are two ways to measure the diameter of the crater. The first is the diameter at the original target surface, and the second is the rim diameter, the diameter across the top of the crater rim. Here we use that rim diameter.

The second important measure is the ratio of the rim diameter to impactor diameter. That ratio provides the link between observations of crater distributions on a solar system body and the impactor flux that created it. In the earliest applications, e.g. Wetherill, 1967, it was simply assumed that that ratio was a single constant; based on a single experiment and the assumption of crater energy-scaling. Even some recent studies have used that same simple assumption. However, we find here that it can be much more complicated than that. The scaling ratio can depend strongly on both the target material, and on the size of the crater; as will be demonstrated.

We present results for each of the three cratering regimes in order. To do so, we assume generic parameter values as follows:

| Property | Value (cgs) | Value (SI) |
|---|---|---|
| Mass density | 2.7 g/cm$^3$ | 2700 kg/m$^3$ |
| Lab tensile Strength | 1.5 10$^8$ dynes/cm$^2$ | 15 MPa |
| Lab Specimen Size | 10 cm | 0.1 m |
| Strength-Size exponent for shear strength | n=4 | |

---

[3] Consistent with that, the number of several cm fragments that would fit on the approximately 30 cm circumference of the crater rim would be about 10.

| | | |
|---|---|---|
| Strength-Size exponent for tensile strength | n=2 | |
| Impact Velocity | 5. 10⁵ cm/s | 5 10³ m/s |
| Impact Angle | 45° | |
| Gravity regime constant $K_1$ in Eq. 23. | 0.84 | |
| Strength Regime constant $K_{DB}$ in Eq. 5. | 1.71 | |
| Spall regime constant $K_S$ in Eq. (9) | 9.34 | |

Table 2.  Generic Properties used for numerical examples.

### 4.1. Cratering in the gravity regime

The gravity regime applies for the largest craters on an asteroid. It applies for craters larger than the transition between the gravity and the strength regimes. That transition size was listed above in Eq. 7. The gravity is related to the target diameter $D_T$ as $g = 2\pi/3\, \rho G D_T$, so that the gravity regime is when

$$\frac{D_{crater}}{D_0} > \left[\frac{3Y_0}{2\pi D_0 D_T G \rho^2}\right]^{\frac{n}{1+n}}. \tag{20}$$

Using the values in Table 2 and converting to km, this transition is

$$(D_{str-grav})_{km} = 342.\,(D_T)_{km}^{-0.8}, \tag{21}$$

As an example, for a 100 km asteroid with the values in Table 2, the strength-gravity limit has the crater diameter of 8.6 km.

The primary scaling formula for a crater size in the gravity regime is given in terms of the non-dimensional forms $\pi_V = \frac{\rho V}{m}$, $\pi_2 = \left(\frac{ga}{U_n^2}\right)$ (e.g. Holsapple, 1993) as

$$\pi_V \sim (\pi_2)^{\frac{-3\mu}{2+\mu}}. \tag{22}$$

For an impact at angle $\phi$ from the vertical, $U_n = U Cos(\phi)$ is the normal component of the impact velocity[4]. The impactor radius $a=d/2$, where $d$ is the impactor diameter. For many materials, $K_1$ is on the order of 0.6. The cube root of Eq. 22 (ignoring a small power of the mass density ratio) results in

$$\frac{D_{crater}}{d} = K_1 \left(\frac{gd}{2U_n^2}\right)^{\frac{-\mu}{2+\mu}}. \tag{23}$$

Again using the expression for the gravity,

$$\frac{D_{crater}}{d} = K_1 \left(\frac{\pi G \rho D_T d}{3 U_n^2}\right)^{\frac{-\mu}{2+\mu}}. \tag{24}$$

Using the values from the Table 2 and converting to km units gives

$$D_{crater} = 40.9\,(D_T)^{-.216} d^{0.784}, \tag{25}$$

As the governing crater size scaling.

The diameter scaling ratio is in Eq. 23, but it is more useful as a function of the crater diameter rather than the impactor diameter. One can solve Eq. 25 for the impactor diameter in terms of the crater diameter and use that expression in the right-hand side of Eq. 23 to obtain this ratio in terms of the crater diameter only:

---

[4] Except the papers at that time did not consider angled impacts.

$$ratio = \frac{D_{crater}}{d} = (K_1)^{(\frac{\mu+2}{2})} \left(\frac{\rho G D_{crater} D_T}{U_n^2}\right)^{-\mu/2}. \tag{26}$$

For a given target, the dominant dependence here is on the crater and impactor sizes. Again, using the constants of Table 2, this diameter scaling ratio 'ratio' decreases with the crater diameter in km as

$$ratio = \frac{D_{crater}}{d_{impactor}} = 54.\,((D_T)_{km}(D_{crater})_{km})^{-0.275}. \tag{27}$$

For an impactor of 10 km diameter (in the gravity regime) at 45° and 5 km/s, the diameter scaling ratio is about 17. From Eq. (27) that value scales with the crater or the target diameter to the power exponent of -0.275.

### 4.2. Cratering in the strength regime

The strength regime is for crater diameters below that of Eq. 21 down to the strength-spall transition. That strength-spall transition crater diameter, using the spall constants listed, is given by Eq. 18. Again, using the relation between the gravity and the target diameter $D_T$, the transition diameter becomes

$$D_S^* = 0.047\, D_0\, (Y_0/(D_0 \rho^2 G D_T))^{2/3}. \tag{28}$$

Using the values in Table 2, and km units, this transition is given by

$$(D_{sp-str})_{km} = 2.15 (D_T)_{km}^{-2/3} \tag{29}$$

As an example, on a Gaspra-sized asteroid with $D_{target} = 12.2\ km$, this transition at about $D$=0.41 km.

For the crater size, starting with Eq. 5, these constants give the bowl crater diameters in meters in the strength regime in terms of the impactor diameter in meters as

$$(D_B)_m = 9.4\,(d_m)^{1.074}. \tag{30}$$

The diameter scaling ratio in the strength regime was given in (9). Again, the dependence on the impactor radius $a$ can be eliminated in favor of the crater diameter to obtain

$$ratio = \frac{D_{crater}}{d_{impactor}} = \frac{K_{DB}}{2}\left(\frac{D_{crater}}{K_{DB}\,D_0}\right)^{\frac{\mu}{2n_1}} \left(\frac{Y_0}{\rho U_n^2}\right)^{-\mu/2}. \tag{31}$$

Thus, this strength scaling ratio increases as $(D_{crater})^{\frac{\mu}{2n_1}}$. Since it is in the strength regime, it does not depend on the gravity or target size.

Using the parameters in Table 2, in the strength regime the crater scaling ratio becomes

$$ratio = \frac{D_{crater}}{d_{impactor}} = 8.05\,(D_{crater})^{0.069}. \tag{32}$$

### 4.3. Cratering in the spall regime

The spall regime is below the crater diameter of Eq. 16. Inserting the expression for the gravity and the generic constants and converting to km provides the largest spall crater as

$$(D_S^*)_{km} = 2.14\,(D_t)_{km}^{-\frac{2}{3}}. \tag{33}$$

The crater diameter scaling in the spall regime is from Eq. 9, which becomes using meter units and the impactor diameter,

$$D_m = 70.\,(d_m)^{1.16}. \tag{34}$$

For the same impactor dimension, this is a factor of several larger than a strength bowl crater as given in Eq. 30.

Finally, the scaling ratio in terms of the crater diameter in the spall regime is

$$ratio = \frac{D_{crater}}{d} = \frac{K_S}{2}\left(\frac{D_{crater}}{K_S D_0}\right)^{\frac{\mu}{2n_2}} \left(\frac{Y_0}{\rho U_n^2}\right)^{-\mu/2}, \qquad (35)$$

which again depends only on strength, crater size and impact velocity. Using the parameter values listed in Table 2, this becomes

$$ratio = \frac{D_{crater}}{d} = 102.\,(D_{crater})_{km}^{0.14}, \qquad (36)$$

for the spall crater regime. These ratios are distinctly larger than bowl craters, a consequence of their broad, shallow shapes.

For an asteroid, using again the expression for the gravity, the boundary for ejecta loss, using km units, is for spall craters smaller than

$$(D_S)_{km} = 0.89(D_T)_{km}^{-4} \qquad (37)$$

### 4.4. Summary of regime ranges

All these results can be plotted on a single figure. The Fig. 8 shows the general regimes of cratering as a function of a target surface gravity (lower scale) or target diameter (upper scale). In addition, the crater type ranges are depicted for two different choices of tensile strength: the red curves are for 15 MPa and the blue curves for 150 MPa. The range for retention of spall fragments is also included. For the asteroids noted on the Fig. 8, most fragment blocks would be retained. Rocky asteroids smaller than 2 km would only have spall craters, but the spall fragments would not be retained.

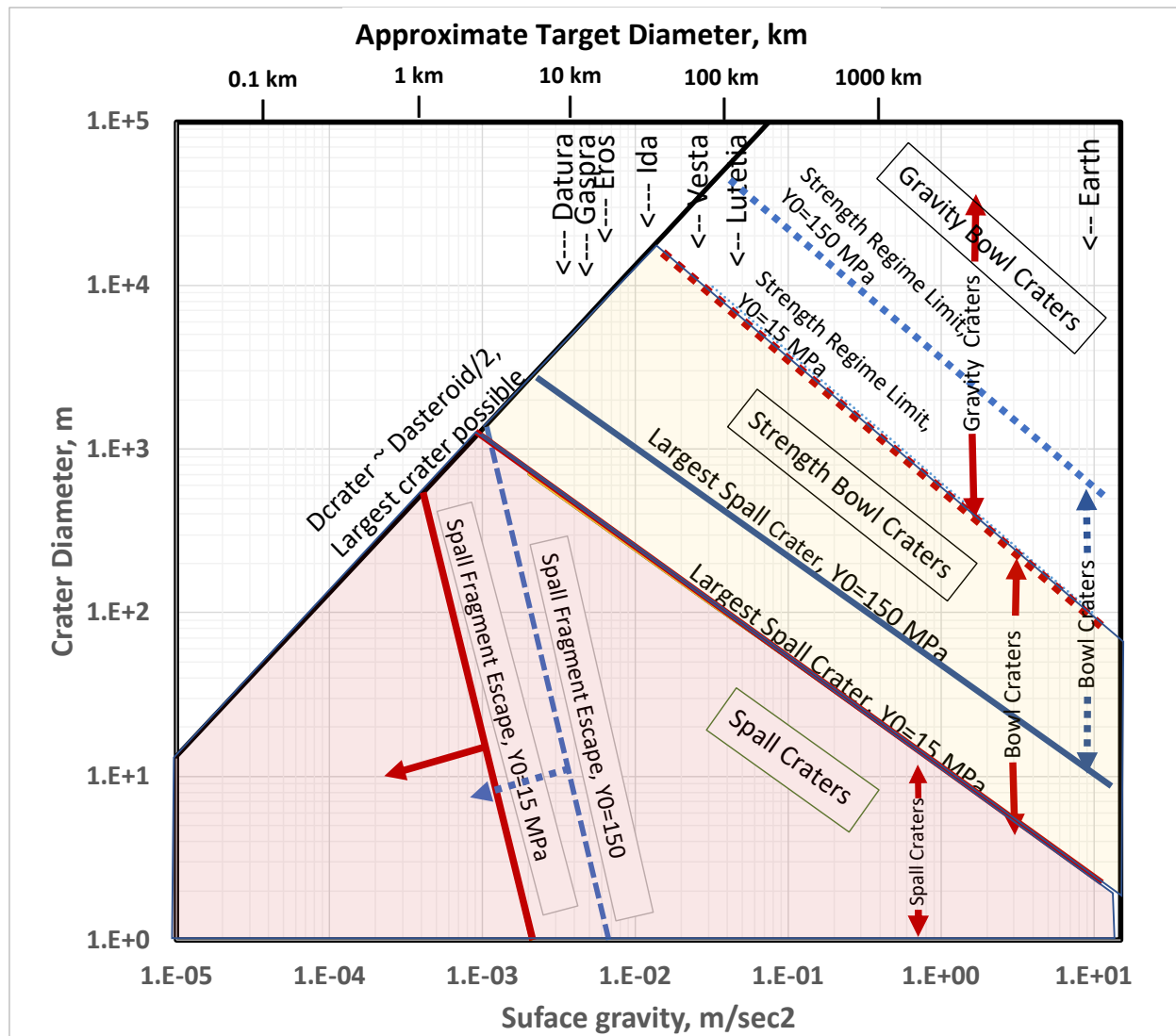

Figure 8. The ranges of possible spall craters (red shaded region), strength bowl craters (light yellow), gravity bowl craters (above dotted red line). The maximum crater sizes without causing disruption and dispersion for these finite objects is on the order of the target radius (but maybe a little larger), is also plotted along the top left.  The range for escaping spall fragments are at the lower left.  The corresponding blue lines are for a tensile strength of 150 MPa.  The diameters of several notable asteroids are indicated at the top.

This shows that for the strongest rocky or icy asteroids smaller than 10 km, all craters are likely to be spall craters.  However, if smaller than 1 km, their spall fragments will not be retained.  For weaker bodies, the largest spall craters might be as large as 3 km, although most will be smaller.

### 4.5. Crater diameter-scaling ratio

As stated above, it is the ratio of the crater diameter to the impactor diameter that is the key to understanding observations of the craters on the asteroids, especially the smaller ones. The various formulas above provide meaningful plots of this scaling ratio versus the crater dimension on an asteroid with a given diameter and assumed properties.

The interested reader should be aware that the formulae here, in addition to other crater scaling formulas for the various cratering regimes, are evaluated numerically and plotted in an online tool (Holsapple, 2022) accessible on any browser at Impacts! There the various parameters are input as desired.  Many results and plots are then available. As an example, using the parameters in Table 2, a plot of the crater scaling ratios for both a 5 km and a 500 km asteroid is obtained as follows:

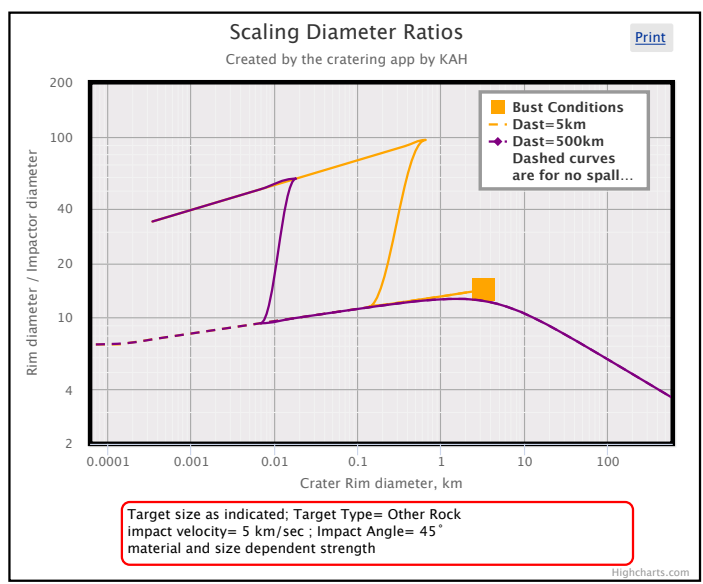

Figure 9.  The crater diameter scaling ratios as a function of the crater diameter for a high strength brittle asteroid for two different asteroid diameters, 5 km and 500 km using rocky impactors.  The material values from Table 2 are used. At the far right, the decreasing slope is in the gravity regime is from Eq. 27. In the central region, with crater diameters from 0.1 to 1 km, the ratio increases with crater diameter according to the Eq. 32. The target size does not matter. Finally, the smallest craters to the left are spall craters and the scaling ratios jump upward by a factor of several.  There is a small range for crater diameters where this curve is double valued. The upper branch is for spall craters and the lower branch is without spall.  The dashed line continues the strength regime to smaller crater diameters, assuming that spall is not a factor.

This figure defines the range of results, from a small 5 km asteroid to a large 500 km one. There are two differences as a result of the 100x change in surface gravity. First, on the larger asteroid, the transition from spall to bowl-strength cratering happens at a crater diameter of around 10 m.  But, on the smaller body, spall craters continue to form up to a diameter of about 300 m because the relatively lower gravity allows spall plates to be ejected from the spall crater.  Notice that in the bowl-strength regime where the diameter ratio is about 10-15, the two curves coalesce because the ratio only depends on material strength, which was assumed to be the same for both asteroids.  The second difference is that, craters larger than several km form in the gravity regime whereas the smaller body has no gravity-dominated craters. The 5 km asteroid would be destroyed when creating a crater with a diameter over a few kilometers. This limit typically occurs for a crater diameter approximately equal to the asteroid radius.

This plot is for relatively strong rocky asteroids. For a small strength C-Type object, the plot is much simpler:

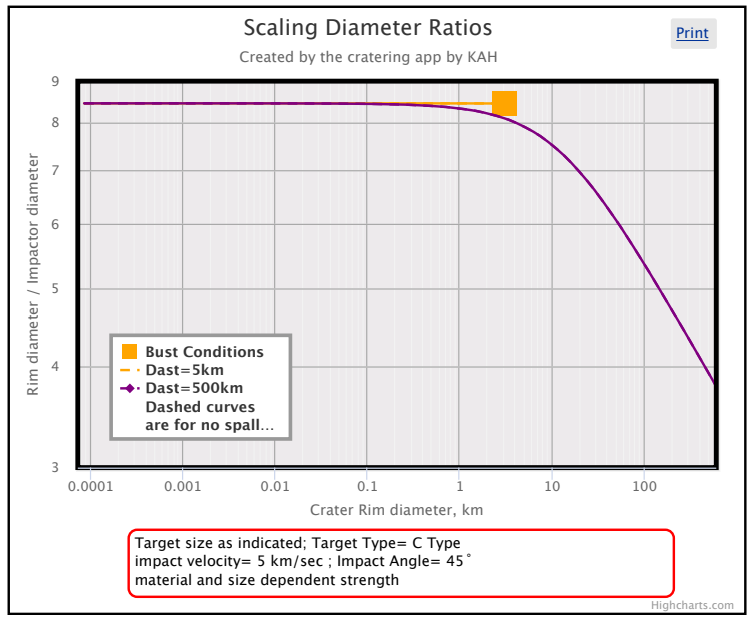

Figure 10. The scaling ratio for C-Type asteroids 5 km and 500 km diameter, and C-Type impactors. Spall craters do not occur. In the strength regime, below about a 1 km crater, the ratio is nearly constant at 8.5  For the 500 km object there is a distinct fall-off in the gravity regime on the right for km-sized craters.

## 4.6. Crater Counts

A major observable of small asteroids is their surface crater counts: the size-frequency distribution (SFD) of the crater sizes. Those counts are a direct outcome of two factors: the SFD of the impactor population, and the mapping between the impactor size and the crater size, i.e. the crater diameter scaling just presented. One can infer the impactor population for an asteroid by assuming a crater scaling relation and using the observed surface crater count. That can be the primary way to estimate small impactor statistics in the main belt. Or one can assume the population distribution and predict the cratering and perhaps use that to deduce surface properties.

This study illuminates possible shortcomings in previous studies that relied on crater scaling relations. It is common to assume that the scaling relation is a simple single constant, on the order of 10, i.e. the diameter of the crater is 10 times that of the impactor . However, as the results here show, that is not generally true. The ratio may be very different for different asteroid types, and even for different crater sizes on the same asteroid. And especially for rocky asteroids, one must consider the possibility of spall craters. The choice of the assumed cratering type will change the relation between the impactor flux and an observed crater count.

It is common to express the cumulative number of impactors with a radius greater than some value $a$ as a power law

$$N_{>a} \sim a^{-q}, \tag{38}$$

with the differential distribution

$$dN_{>a} \sim a^{-p} da. \tag{39}$$

The exponent $p=q+1$ is called the power law index of the differential distribution.

Assuming for now simple bowl craters in the strength regime, the Eq. 5 has

$$D_B \sim a^{2n/(2n-\mu)}. \tag{40}$$

Solving for the impactor radius *a,* and using that in Eq. 38 gives the resulting crater size distribution as

$$N_{>D_B} \sim D_B^{-q(2n-\mu)/(2n)}. \tag{41}$$

If one obtained this exponent from observed crater counts, then one could solve for the power-law distribution of impactor sizes. Or if the impactor index is provided then this determines a distribution of the impact crater sizes.

Consider Gaspra as an example. Chapman et al., 1996 report a differential crater population index of 4.3± which gives a cumulative crater distribution power of 3.3. If we denote the exponent in (41), as *s,* the impactor population has an index of $q = \frac{2ns}{2n-0.55}$ where $\mu$=0.55. In the special case where the asteroid strength is not size-dependent, then n -> ∞ and *q*=3.3, the same as the crater distribution exponent *s*. However, if we consider the more physically reasonable value of *n*=4, then *q*=3.54. For an even stronger size dependence with *n*=2 the inferred impactor population index of 3.8. Therefore, *a size-dependent strength results in the exponent for the impactor size distribution having a larger magnitude (steeper slope) that that for the crater population*.

An even larger effect occurs if the actual craters are formed by spall and not simple shearing and ejection of material. Additionally, it could be that only the smaller diameter part of the range of observed craters are spall craters.

First suppose that all the craters form in the spall regime. Assuming a size-dependent strength and a given target body, the crater scaling was given above in Eq. (9) as

$$D_S \sim (a)^{\frac{2n_2}{2n_2-\mu}}. \tag{42}$$

where the exponent $n_2$ defines the size dependence of the strength for spall craters. Just as shown above, the size dependent strength causes the causes the power-law exponent of crater size to be larger than that of the impactor population. But that is not the biggest effect. It is the much larger diameters that result from spall craters for a given impactor that make the primary difference.

Consider the crater diameters corresponding to some bin of impactor diameters. If the craters are bowl shaped, then using (41) the cumulative number count of those craters will fall along some curve in the log-log plot of the cumulative crater size count. An idealized example is shown in Fig. 11. However, if a crater was formed by spall, the crater diameter will be a factor of several larger than the bowl crater, and the data point will be at a much larger diameter at that same number *N*. From that diameter and all smaller ones, all points will be offset to the right in Fig. 11 by a factor of several. Equally well, being a power law, at a given diameter it will be offset up to a larger number by a factor of several times the impactor index.

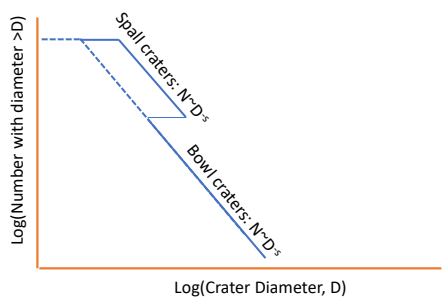

Figure 11. The nature of the cumulative crater size distribution when there is a transition from bowl shaped craters to spall craters in the strength regime. All larger craters are bowl craters, but at some value there is a transition to spall cratering. The curve for the spall crater data will be offset to the right by a factor of several, depending upon the material parameters of the asteroid. In fact, one would not expect the transition to be that sharp so there would be a fuzzy area around the transition diameter.

If one had discrete points along this broken curve, they might interpret that as a steeper slope for a crater size distribution function. And then the inferred impactor distribution would also be too steep. Conversely, if one used an assumed impactor SFD and a constant crater scaling ratio, one might be puzzled by the lack of small craters, because the diameter of the otherwise small bowl craters would be enhanced by spall. There would be a perceived "shortage" of small craters. Instead, there would be increased number of blocks, which would be a characteristic of surface spall craters. Broad flat spall craters could be interpreted as degraded craters. Eject blankets would be absent. In fact, those features have been noted on several small asteroids (e.g. Chapman et al., 1996).

## 5. Summary

On Earth, experimental cm-sized craters in brittle materials are always spall craters as described, while 10 m craters are not. We find here that the spall crater formation mechanism is limited to craters below a size determined by the surface gravity and the surface tensile strength. That is very important regarding our interpretations of cratering on small asteroids.

The fact that spall craters are limited in size is likely a result of two factors. They arise when the shockwave from the impact is reflected from the 'top' free surface as a tensile wave, that tension exceeds the strength of the material, and a plate with a thickness corresponding to the width of the tensile pulse is lifted from that top surface. For larger impactor and crater size, the spall plates are correspondingly thicker, but are launched with the same vertical velocity at all size scales. That implies a fixed launch height. However, to be removed from their initial locations they must be launched to a height at least comparable to their thickness. Therefore, for larger events, the spall plates are not launched with sufficient velocity to leave their initial location. According to this explanation, the Polanskey and Ahrens, 1990 data for spall thicknesses and velocities from Gabbro craters with typical diameters of 6-10 cm would imply the absence of the external spall region for craters on Earth larger than about 5 m. That estimate are consistent with the field data for explosion craters (Gault, 1973).

The lower gravity on an asteroid accentuates these factors. On a km-sized asteroid consisting of a brittle rock material, all craters can be expected to be of the spall type. And for a 15 km asteroid, craters with a diameter of a km or larger would not be spall craters, but the smaller ones would be. The spalling of the 100 m sizes would create blocks with dimensions on the order of 10s of meters. The km sized and larger craters would have a fractured zone, of the order 3-4 times their diameter surrounding them, but they would not launch discrete blocks.

Here we have used those ideas to formulate a physical framework that governs spall cratering. To date these concepts have not been applied to asteroids but they may be very important for the observations of small rocky asteroids. There is much to be studied and learned. It will ultimately be up to experimenters and observers to test the conjectures and refine the theory.

## 7. Author contributions.

KAH was responsible for developing the theory, writing this manuscript, and collecting the resources presented. KRH noted the fundamental mechanisms of spall cratering described in the Gault et al, reference, and assisted in the writing of the manuscript.